\documentclass{article}

\usepackage{arxiv}

\usepackage[utf8]{inputenc} 
\usepackage[T1]{fontenc}    
\usepackage{hyperref}       
\usepackage{url}            
\usepackage{booktabs}       
\usepackage{amsfonts}       
\usepackage{nicefrac}       
\usepackage{microtype}      
\usepackage{lipsum}
\usepackage{amsmath}
\usepackage{graphicx}
\usepackage{color}

\DeclareMathOperator*{\argmin}{arg\,min}

\newcommand{\bsb}{\mbox{\boldmath $\beta$}}

\newcommand{\mbx}{\mathbf{x}}

\begin{document}

\title{Bayesian Ordinal Quantile Regression with a Partially Collapsed Gibbs Sampler}

\author{
  Isabella N. Grabski \\
  Department of Biostatistics\\
  Harvard University\\
   \And
 Roberta De Vito \\
  Department of Computer Science\\
  Princeton University\\
  \And 
  Barbara E. Engelhardt \\
  Department of Computer Science \\
  Princeton University
}

\maketitle

\begin{abstract}
{Unlike standard linear regression, quantile regression captures the relationship between covariates and the conditional response distribution as a whole, rather than only the relationship between covariates and the expected value of the conditional response. However, while there are well-established quantile regression methods for continuous variables and some forms of discrete data, there is no widely accepted method for ordinal variables, despite their importance in many medical contexts. In this work, we describe two existing ordinal quantile regression methods and demonstrate their weaknesses. We then propose a new method, Bayesian ordinal quantile regression with a partially collapsed Gibbs sampler (BORPS). We show superior results using BORPS versus existing methods on an extensive set of simulations. We further illustrate the benefits of our method by applying BORPS to the Fragile Families and Child Wellbeing Study data to tease apart associations with early puberty among both genders. Software is available at: {\tt GitHub.com/igrabski/borps}.}
\end{abstract}

\section{Introduction}
Standard linear regression captures relationships between covariates of interest and the conditional mean of the response, modeled as a linear function. In particular, the linear regression model is based on the expected value of our response variable $y$ given the values of the $p$ covariates $\mbx \in \mathcal{R}^p$. However, this model is unable to capture relationships between the covariates and the conditional distribution $y|\mbx$ as a whole, as it is limited instead to a linear model of the expected value with Gaussian residuals. 

Quantile regression \cite{koenker1978regression} allows us to quantify a more complex relationship between the covariates and the distribution of the response variable by modeling the conditional quantile function of response variable $y$, $Q_{y}(q|\mbx)$, with $0<q<1$.  Here, the $q$th quantile of $y$ given covariates $\mbx$ is defined as $\inf \{y : F(y) \geq q \}$ for the cumulative distribution function $F(y)$. As a result, quantile regression estimates both the variable effects of covariates across conditional quantiles and the shape of response distributions conditional on $\mbx$. This is useful in any situation where the mean might not adequately describe the conditional response distribution, such as in the presence of non-Gaussian residuals. Quantile regression allows us to uncover interesting structure that might be present in the tails of the distribution, including heavy-tailed or skewed distributions, that would otherwise be masked in standard regression and distort inference.

While there are well-established quantile regression methods for continuous outcomes in both frequentist \cite{koenker2001quantile} and Bayesian \cite{yu2001bayesian} frameworks, there are fewer methods for discrete outcomes. In particular, there is no widely accepted quantile regression method for ordinal variables, i.e., ordered variables without an underlying interval. Ordinal variables are especially common in medical contexts, where many health outcomes are expressed as ordered categories rather than as strictly numerical measures, (e.g., stages of cancer, BMI categories, or grades of disease severity). However, there are also numerous examples of ordinal variables in finance (e.g., corporate credit ratings \cite{hirk2018multivariate}), ecology (e.g., organism maturity \cite{deyoreo2018modeling}), transportation (e.g., transportation quality metrics \cite{allen2018effect}), and many other disciplines.

The standard approach to regression with ordinal response variables is to use the ordinal probit model. In this model, each ordinal variable is modeled by an underlying continuous latent variable, and these latent variables are linearly related to the covariates with Gaussian residuals. Just as with standard linear regression, fitting this ordinal probit model only captures the mean of the conditional distribution of the continuous latent variable underlying each response. In order to study the full conditional distributions of such outcomes, rather than assuming Gaussianity and modeling the empirical means, quantile regression methods may be used for the analysis of ordinal variables. 

Two recent ordinal quantile regression methods have been described in the Bayesian framework \cite{rahman2016bayesian,alhamzawi2017bayesian}; we will show that both are problematic in their current forms, because of sensitivity to parameter specifications in the former case, and poor performance on noisy simulations in the latter. 
To fill this gap, we propose a new method, \underline{B}ayesian \underline{O}rdinal Quantile \underline{R}egression with a \underline{P}artially Collapsed Gibbs \underline{S}ampler (BORPS). We demonstrate superior results to existing methods on an extensive set of simulations. We further demonstrate the utility of our method by applying BORPS to a large data set to study the relationship of early puberty to other childhood physiological and demographic factors.

\label{sec1}

\section{Methods}
\label{sec2}

We denote our observed response variables $y_i$, for $n$ samples indexed by $i \in \{1,\dots,n\}$, with $p$ corresponding covariates $\boldsymbol{x_i} = \big[ x_{i,1}, \ldots, x_{i,p} \big]$. 
We begin by reviewing quantile regression for continuous $y_i \in \Re$ in the classical approach and then in a Bayesian framework.
Then, we discuss existing quantile regression approaches for ordinal $y_i$. Finally, we describe our quantile regression method for ordinal $y_i$ and associated inference methods.

\subsection{Quantile regression}

In quantile regression, we are interested in estimating the coefficients $\hat{\bsb}$ of the $q$th quantile of $\boldsymbol{y}|\boldsymbol{x}$, denoted $Q_{Y|\boldsymbol{x}}(q)$, in the model $Q_{Y|\boldsymbol{x}}(q) = \bsb \boldsymbol{x}$. 
Consider the usual definition of quantiles, where the cumulative distribution function of random variable $y$ is written as $F_Y = P(Y \leq y)$. For $q \in (0,1)$, the $q$th quantile of $Y$ is the real value $Q_{Y}(q)=F_{Y}^{-1}(q) =\inf \left\{y : F_{Y}(y)\geq q \right\}$.  
In the context of regression, we include covariates in this definition, and we model the cumulative distribution function of this conditional distribution as $F_{Y|\boldsymbol{x}}(y) = P(Y \leq y|\boldsymbol{x})$. Thus, we are interested in $\hat{\bsb}$ at the $q$th quantile of $Y|\boldsymbol{x}$, i.e., $Q_{Y|\boldsymbol{x}}(q) = \inf \{ y : F_{Y|\boldsymbol{x}}(y) \geq q \}$.

In the classical approach, inference in quantile regression is formulated as an optimization problem for a linear model, analogous to the ordinary least squares (OLS) optimization used for standard linear regression \cite{koenker2001quantile}. OLS for linear regression finds the regression coefficients $\hat{\bsb}$ that minimize the sum of squared residuals
 \begin{equation} 
 \hat{\bsb} = \argmin_{\beta \in {\rm I\!R}^p} \sum_{i=1}^n (y_i - \mbx_i'\bsb)^2. 
 \end{equation} 
This yields the optimal coefficients $\bsb$ to model the conditional mean.
 
Inference in quantile regression takes a similar approach, but since we are now interested in any $q$th conditional quantile the optimization problem depends on $q$. To do this, we introduce the \emph{check function} $\rho_q(\cdot)$, which asymmetrically weights positive and negative residuals by the quantiles $q$ and $1-q$. The associated optimization problem can be written as
 \begin{equation} 
 \hat{\bsb} = \argmin_{\beta \in {\rm I\!R}^p} \sum_{i=1}^n \rho_q(y_i - \mbx_i'\bsb).
 \end{equation} 
Formally, the check function for arbitrary random variable $u$ is defined as $\rho_q(u) = u(q - {\rm I\!I}(u < 0))$, or equivalently, $\rho_q(u) = q|u|{\rm I\!I}(u \geq 0) + (1-q)|u|{\rm I\!I}(u < 0)$, where $\rm I\!I$ denotes the indicator function. Hence, the optimization problem for quantile regression can be written as 
\begin{equation} 
\hat{\bsb} = \argmin_{\beta \in {\rm I\!R}^p} \left( \sum_{i:y_i \geq x_i'\beta} q|y_i - \mbx_i'\bsb| + \sum_{i:y_i < x_i'\beta} (1-q)|y_i - \mbx_i'\bsb| \right). 
\end{equation}

Note that, at $q = 0.5$, the optimization problem reduces to 
\begin{equation}
    \hat{\bsb} = \argmin_{\beta \in {\rm I\!R}^p} \sum_{i=1}^n 0.5|y_i - \mbx_i'\bsb| = \argmin_{\beta \in {\rm I\!R}^p} \sum_{i=1}^n |y_i - \mbx_i'\bsb|,
\end{equation}
which can be recognized as standard median regression. 

This approach to quantile regression received a great deal of traction when first introduced and is still widely used today \cite{davino2013quantile}.
There was minimal development of Bayesian approaches to quantile regression until more recently \cite{yu2001bayesian}. 
Like the classical approach, the Bayesian version of quantile regression adopts the check function. But the Bayesian approach relies on the observation that the asymmetric Laplace distribution ($\mathcal{ALD}$) contains a check function within its probability density function (PDF). Indeed, if $u \sim \mathcal{ALD}(\mu,\sigma,q)$ for location $\mu$, scale $\sigma,$ and skew $q$, then the $\mathcal{ALD}$ PDF is written as: 

\begin{equation} 
f(u| \mu, \sigma, q) = \frac{q(1-q)}{\sigma}\exp \left( -\rho_q\left(\frac{u - \mu}{\sigma}\right) \right),
\end{equation}
for check function $\rho_q(\cdot)$.

In this framework, when adopting an $\mathcal{ALD}$ prior on the residuals $\epsilon_i$ with $\mu = 0$ and $\sigma = 1$, the likelihood function of our model $y_i = \mbx_i'\bsb + \epsilon_i$ becomes \begin{eqnarray*} 
L(y|\beta) & = & q^n(1-q)^n \exp \left(-\sum_{i=1}^n \rho_q(\epsilon_i) \right) \\
&=& q^n (1-q)^n \exp \left(-\sum_{i=1}^n \rho_q (y_i - x_i'\beta) \right). 
\end{eqnarray*}
The likelihood function when $\epsilon \sim \mathcal{ALD}(0,1,q)$ can then be seen to contain the objective function for the $q$th quantile from the classical approach. 
By developing an appropriate Markov chain Monte Carlo (MCMC) method, we can perform Bayesian inference to estimate the full posterior distribution $\bsb | 0,1,q,\boldsymbol{X},\boldsymbol{y}$; we can select point estimates $\hat{\bsb}$ at any desired quantile by querying the posterior with the skew parameter $q$ set to that quantile.

Previous approaches developed sampling methods for Bayesian quantile regression. Prior work used a random walk Metropolis-Hastings method~\cite{yu2001bayesian}. Other methods that are more computationally efficient and require less parameter tuning, including a Gibbs sampler \cite{kozumi2011gibbs} and a partially collapsed Gibbs sampler \cite{reed2009partially}, have also been developed. Nonparametric Bayesian quantile regression methods take a completely different approach than the work described here, by either leveraging the Bayesian exponentially tilted empirical likelihood \cite{lancaster2010bayesian} or by using nonparametric distributions to model either the error distribution \cite{reich2009flexible} or the joint distribution of the response and covariates \cite{taddy2010bayesian}.

The Bayesian quantile regression methods discussed here have thus far been developed under the assumption that the response variable $y$ is continuous. There are well-established frequentist quantile regression methods to handle binary \cite{bottai2010logistic} and count \cite{machado2005quantiles} response variables. Some of these approaches have Bayesian alternatives \cite{benoit2012binary,lee2010bayesian}. Work on ordinal variables in the frequentist statistical framework has been more recent~\cite{hong2013multi,hong2010prediction}. Here, we focus our attention on Bayesian approaches, which allows us to take a more flexible modeling approach through our choice of prior distributions. 

\subsection{Bayesian Quantile Regression for Ordinal Response Variables}
Two Bayesian approaches for quantile regression with ordinal response variables have recently emerged, which we will refer to as Alhamzawi's Method~\cite{alhamzawi2017bayesian}, or AM, and Rahman's Method~\cite{rahman2016bayesian}, or RM, respectively. Both are Bayesian methods for ordinal quantile regression based on similar frameworks, differing largely in their choice of parameterization and priors. 

AM was originally developed for longitudinal ordinal variables. Here, we consider the model in the simplest case, i.e., a single time point. This method is similar to a Bayesian approach for Gaussian quantile regression~\cite{yu2001bayesian}, with a key difference in the treatment of the response variable $y$. Since $y$ is continuous in Gaussian quantile regression, its residuals may be modeled directly as $\mathcal{ALD}$ variables; this direct approach is not meaningful for ordinal $y$. To handle this, AM introduces continuous latent variables $z_i$ corresponding to each $y_i$, as well as a global cutpoint vector $\delta$. Suppose $y$ can take on $C$ possible ordered values, which we encode as $y \in \{ c_1, c_2, \ldots, c_C\}$. Then these parameters are related as follows: 
\begin{equation} 
y_i = \begin{cases} c_1 \text{ if } \delta_0 \leq z_i < \delta_1 \\ c_j \text{ if } \delta_{j-1} \leq z_i < \delta_j; \;j = 2, \ldots, C-1 \\ C \text{ if } \delta_{C-1} \leq z_i < \delta_C \end{cases}. 
\label{eq:yz}
\end{equation} 

This formulation allows us to transform the ordinal response into a continuous one. Whereas \cite{yu2001bayesian} models $y = \mbx'\bsb + \epsilon$ for continuous and observed $y$, AM models $z = \mbx'\bsb + \epsilon$ for continuous but unobserved $z$, which is related to the ordinal response variable $y$ as in \ref{eq:yz}. In both cases, $\epsilon$ is drawn from an $\mathcal{ALD}$. The coefficients have Laplace priors, and the cutpoint vector's prior is the order statistics from a uniform distribution. Gibbs sampling can then be used to perform inference on this model, leveraging the key observation that the $\mathcal{ALD}$ can be written as a conditionally conjugate normal-exponential mixture. Thus, AM estimates the regression quantiles for $z$, which are straightforward, in order to estimate the more challenging regression quantiles for ordinal variables $y$.

RM uses a similar model to AM but includes a different parameterization of the $\mathcal{ALD}$ for Gibbs sampling, as well as some different prior choices~\cite{rahman2016bayesian}. The biggest difference between the two approaches is that RM does not update the values of the global cutpoint vector $\delta$ during Gibbs sampling. Instead, the cutpoint values are set ahead of time and remain fixed throughout inference. This model was termed ``ORII" in \cite{rahman2016bayesian} and is intended for ordinal response variables that have exactly three categories; an alternative model ``ORI" uses a Metropolis-Hastings step to update $\delta$ and can accommodate more than three categories for the response variable. However, we were unable to achieve a reasonable acceptance rate using this model in our simulations, so we only consider ORII in this work.

\subsection{Bayesian Ordinal Quantile Regression with a Partially Collapsed Sampler (BORPS)}
Our method, \underline{B}ayesian \underline{O}rdinal Quantile \underline{R}egression with a \underline{P}artially Collapsed \underline{S}ampler (BORPS), combines elements of AM \cite{alhamzawi2017bayesian} and RM \cite{rahman2016bayesian} with important extensions for robustness. In particular, we use a similar Bayesian formulation of ordinal quantile regression from RM, but we incorporate the $\boldsymbol{\delta}$ sampling step of AM. In addition, while AM and RM each use a variation of Gibbs sampling \cite{kozumi2011gibbs} for inference, we extend the partially collapsed Gibbs sampler presented by \cite{reed2009partially} to ordinal variables. In particular, \cite{reed2009partially} developed a partially collapsed Gibbs sampler for classical Bayesian quantile regression by augmenting the data with latent weights and integrating out those weights. The resulting method has fewer parameters to estimate than the full Gibbs sampler, and appears in practice to mix better by collapsing the posterior space, motivating this choice for ordinal quantile regression.

Consider ordinal responses $y_i$ and corresponding covariates $\boldsymbol{x}_i$, for $i = 1, \ldots, n$. As before, the goal is to estimate the coefficients $\hat{\bsb}$ of the $q$th quantile of $\boldsymbol{y}|\boldsymbol{x}$, denoted $Q_{Y|\boldsymbol{x}}(q)$, in the model $Q_{Y|\boldsymbol{x}}(q) = \bsb \boldsymbol{x}$. 

Following the work of \cite{alhamzawi2017bayesian} and \cite{rahman2016bayesian}, we model these ordinal variables $y_i$ in terms of continuous latent variables $z_i$ as in Equation \ref{eq:yz}, where $C$ is the number of ordered responses each $y_i$ can take on, and $\boldsymbol{\delta} = \big[ \delta_0, \ldots, \delta_C \big]$ is our global cutpoint vector. In words, we can think of these elements of the cutpoint vector as thresholding the continuous $z_i$ into the ordinal values of $y_i$. A subject whose latent value $z_i$ falls into the $c$th interval, as defined by successive elements of the cutpoint vector, will have observed response $y_i$ equal to the $c$th largest possible ordinal value.

We then model $z_i = \boldsymbol{\beta}^T \boldsymbol{x_i} + \epsilon_i$ for coefficient $\boldsymbol{\beta} = \big[ \beta_1, \ldots, \beta_k \big]$ and residual $\epsilon_i \sim \mathcal{ALD}(0,\sigma,q)$, where $0 < q < 1$ is our quantile of interest.

We rewrite our model as $z_i = x_i'\beta + \sigma \theta w_i + \sigma \tau \sqrt{w_i} u_i,$ using the normal-exponential mixture of the $\mathcal{ALD}$ as in \cite{rahman2016bayesian}. Here, $w_i \sim \mathcal{E}(1)$, $u_i \sim \mathcal{N}(0,1)$, $\theta = \frac{1-2q}{q(1-q)}$, and $\tau = \sqrt{\frac{2}{q(1-q)}}$. This form of the model allows us to develop a Gibbs sampler, because the conditional distribution of $z_i$ can be written in closed form as $z_i \sim \mathcal{N}(x_i'\beta + \theta v_i, \tau^2 \sigma v_i)$ for $v_i = \sigma w_i$. As in prior work \cite{rahman2016bayesian,reed2009partially}, we place a gamma prior on $\sigma^{-1}| c_0, d_0 \sim \Gamma(c_0,d_0)$ and a multivariate normal prior on $\beta | \boldsymbol{b_0},B_0 \sim \mathcal{MVN}(\boldsymbol{b_0},B_0)$. Following \cite{alhamzawi2017bayesian}, we use an order statistic from the uniform distribution for $\boldsymbol{\delta}$, i.e., $P(\delta) = (C-1)! \left( \frac{1}{\delta_C - \delta_0} \right) \mathbb{I}(\delta \in T)$ when $T = \{ (\delta_0, \ldots, \delta_C) | \delta_0 < \cdots < \delta_C \}$. In other words, each $\delta_j$ is drawn from a uniform distribution over the \textit{a priori} fixed interval $(\delta_0, \delta_C)$, subject to the condition that $\delta_j$ lies between $\delta_{j-1}$ and $\delta_{j+1}$.

We further extend our model by introducing a partially collapsed step in our Gibbs sampler to improve robustness and mixing. In particular, the collapsed Gibbs sampler integrates out $v_i$ from the conditional posterior distribution of $\sigma$. We follow an earlier approach to obtain the collapsed Gibbs sampler in BORPS \cite{reed2009partially}:

\begin{enumerate}
 \item Sample $\sigma^{-1}$ from $\Gamma (c_0 + n, d_0 + \sum_{i=1}^n \rho_q(z_i - x_i'\beta))$.
 \item Sample $v_i^{-1}$, for $i$ ranging from 1 to $n$, from $\mathcal{IG}\left(\frac{1}{q(1-q)|z_i - x_i'\beta|}, \frac{1}{2\sigma q(1-q)}\right)$. Note that here $\mathcal{IG}$ refers to inverse Gaussian, not inverse Gamma. 
 \item Sample $\beta$ as $\mathcal{MVN}\left(\widehat{\beta},\left(\frac{q(1-q)}{2\sigma}x'Vx + B_0^{-1}\right)^{-1}\right)$, for \\ $\widehat{\beta} = \left(\frac{q(1-q)}{2\sigma}x'Vx + B_0^{-1}\right)^{-1}\left(\frac{q(1-q)}{2\sigma}x'Vu + B_0^{-1}b_0\right)$, where $V$ is the diagonal matrix with elements $v_i^{-1}$, and $u$ is the vector with elements $u_i = z_i - \frac{(1-2q)v_i}{q(1-q)}$. 
 \item Sample $z_i$, for $i$ from 1 to $n$, from $\mathcal{TN}_{(\delta_{j-1},\delta_j)}(x_i'\beta + \theta v_i, \tau^2 \sigma v_i)$ when $z_i = j$. 
 \item Sample $\delta_j$, with $j$ from 1 to $C-1$, from a uniform distribution over the interval $(\min \{ \max(z_i | y_i = c), \delta_{c+1}, \delta_C \}, \max \{ \min(z_i | y_i = c + 1), \delta_{c-1}, \delta_0 \} )$. 
\end{enumerate}

We initialize the sampler as follows. For $p$ covariates and $n$ samples, we set $(\beta_1, \ldots, \beta_p) = (0, \ldots, 0),$ and $(z_1, \ldots, z_n) = (1, \ldots, 1)$, $(v_1, \ldots, v_n) = (1, \ldots, 1)$. If $\boldsymbol{R} = \sum_{j=1}^p \boldsymbol{x}_j$, we further set $\delta_1 = 2 \cdot u_1 \cdot \boldsymbol{R}_{0.33}$ and $\delta_2 = 2 \cdot u_2 \cdot \boldsymbol{R}_{0.67}$, subject to the condition that $\delta_2 > \delta_1$, where $u_1, u_2 \sim \mathcal{U}(0,1)$, and we denote the $q$th quantile of $\boldsymbol{R}$ as $\boldsymbol{R}_q$. As hyperparameters, we set $c_0 = d_0 = 10^{-3}$, $\boldsymbol{b_0} = \boldsymbol{0}_p,$ and $B_0 = 10^6 \cdot \mathbb{I}_p,$ where $\mathbb{I}$ denotes the identity matrix.

The derivations for steps 1 and 2 directly follow from \cite{reed2009partially}. Steps 3 and 4 are from \cite{rahman2016bayesian} and step 5 is from \cite{alhamzawi2017bayesian}. See Section 1 of the supplement for full derivations.

BORPS is not identifiable in its current form. As is generally true for ordinal models using this type of latent variable framework, the likelihood can remain the same under both location shifts and re-scaling. Hence, some form of location and scale restrictions are necessary for identifiable parameter estimation \cite{jeliazkov2008fitting}. 
RM tries to enforce identifiability through location fixing, which is why the values of $\boldsymbol{\delta}$ were fixed $\textit{a priori}$. As we will show, fixing $\boldsymbol{\delta}$ leads to poor performance. To allow adaptive cutpoints while still yielding identifiable results, we have two extensions. First, we do not include an intercept term (effectively setting it to 0), which introduces a location restriction without fixing any of the cutpoints. This effectively fixes the mean of the response, which is necessary for identifiability.  
Second, we report $\frac{\boldsymbol{\beta}}{\delta_{C-1}}$, by taking the ratio of their posterior means, instead of $\boldsymbol{\beta}$. This introduces a scale restriction without having to fix the variance of the residuals to a pre-determined value.  
Then, to test statistical significance of the coefficients across quantiles, we use these values in the test statistic to allow us to test whether each $\beta$ at a given quantile is significantly different from 0 or not. 

\begin{figure}
    \centering
    \includegraphics[width=0.45\textwidth]{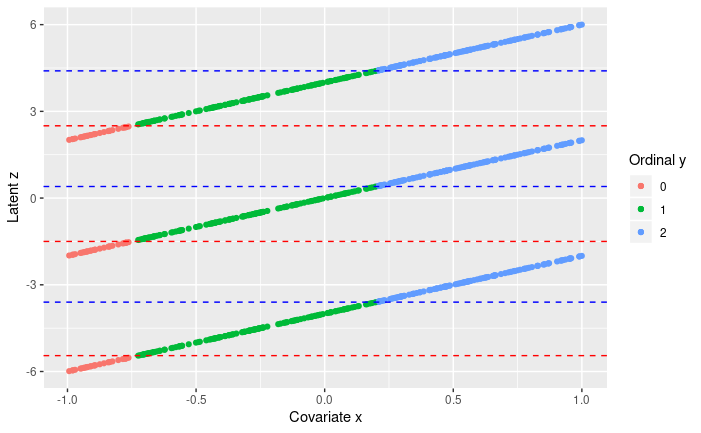}
    \includegraphics[width=0.45\textwidth]{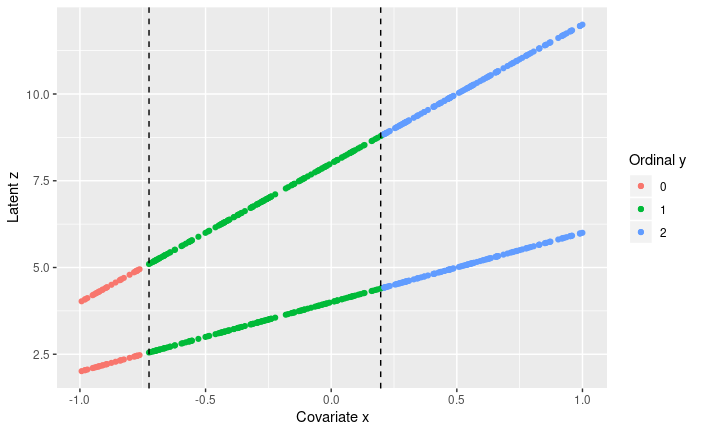}
    \caption{(Left) Shifting the intercept and cutpoint vectors of the model accordingly can result in the same true coefficient for the covariate. The blue dashed line in each case indicates $\delta_2$ (the second value of the cutpoint vector), and the red dashed line indicates $\delta_1$ (the first value of the cutpoint vector). This shows that the model is not identifiable under location shifts. (Right) Scaling the coefficients and cutpoint vector values by the same factor can change the magnitude of the true coefficient, but not which observed ordinal responses are associated with which covariates. The dashed lines indicate the covariates corresponding to each category of observed ordinal response. This shows that the model is not identifiable under scaling shifts.}
    \label{fig:ident}
\end{figure}

We illustrate these identifiability ideas in a simple scenario (Figure~\ref{fig:ident}). First, we show that shifting both the intercept and the cutpoint vectors of the model accordingly results in the same true coefficient for the covariate. This motivates fixing the intercept in order to make the cutpoint vectors identifiable. Second, we show that scaling the coefficients and cutpoint vector values by the same factor changes the magnitude of the true coefficient of the covariate. However, it can be seen that the same ordinal responses are associated with the same covariate values. Hence, inference should be based on the ratios of the coefficients to the cutpoint vector, which are identifiable, rather than on the magnitudes of the coefficients. 

To build confidence intervals for our estimates of $\frac{\boldsymbol{\beta}}{\delta_{C_1}}$, we use bootstrapping. In particular, we repeat $100$ times the following resampling process: We sample the observed sample with replacement, repeat the Gibbs sampling algorithm, and take the desired quantiles of the resulting bootstrapped estimates of $\frac{\boldsymbol{\beta}}{\delta_{C_1}}$ as our confidence interval.

\section{Simulations and Results}

We evaluated results from frequentist continuous quantile regression, AM, RM, and BORPS on a range of simulations, assessing the performance in both univariate and multivariate ordinal quantile regression. We also evaluated a version of AM in which we modified the method to fix the intercept to 0 and report the ratio $\frac{\bsb}{\delta_{C-1}}$, as in BORPS. Our goal is to replicate noisy data in our simulations, and to evaluate performance across these methods in cases where one or more coefficients do not substantially differ from zero. 

In all single-covariate simulations, the simulated data consist of $300$ ordinal responses $y_i$, each with a scalar covariate $x_i$ generated from a uniform distribution on the interval $(0,4)$. The generation of the underlying continuous response $z_i$ varied in each case, but we always set $\delta_1 = 5$ and $\delta_2 = 8$ to threshold each of these $z_i$ into one of three ordinal responses, i.e., $z_i < 5$ yields $y_i = 1,$ $5 \leq z_i < 8$ yields $y_i = 2,$ and $z_i \geq 8$ yields $y_i = 3$. Thus, the observed data in all single-covariate simulations consist of the set of $(y_i, x_i)$ pairs, where $i = \{1,\dots,300\}$. 

For the single-covariate simulations with non-null association, we again generated $z_i$ as $3x_i + u_i$, but used either $u_i \sim \mathcal{N}(\mu_{N,q},1)$ or $u_i \sim \mathcal{L}(\mu_{L,q},1)$, where $\mathcal{L}$ denotes the Laplace distribution. For $q \in \{ 0.25, 0.50, 0.75 \}$, we selected values of $\mu_{N,q}$ and $\mu_{L,q}$ so that the $q$th quantile of $\boldsymbol{u}$ would be 0. This results in six different simulations, where we know the ground-truth coefficient at the given quantile for one of two possible error distributions. 

For the single-covariate simulations with null association, we generated $z_i$ as $12u_i$, 
with either $u_i \sim \mathcal{N}(0,1)$ or $u_i \sim \mathcal{L}(0,1)$. We still use the same covariates $x_i$ described above; hence, these two simulations represent the case of null association because $z_i$ is not a function of the covariates. This implies that the ground-truth coefficient at every quantile should be 0 for both possible distributions of $u_i$. 

In all multiple-covariate simulations, the simulated data consist of 300 ordinal responses with two covariates $x_{1i}$ and $x_{2i}$ generated from a uniform distribution on the interval $(0,4)$ and from a uniform distribution on the interval $(0,2)$ respectively. Again, the underlying continuous response $z_i$ was generated differently in each case, but we thresholded each $z_i$ into one of three ordinal responses $y_i$ with $\delta_1 = 5, \delta_2 = 8$ as we did for the single-covariate simulations. Thus, the observed data across these multiple-covariate simulations consist of the set of $(y_i, x_{1i}, x_{2i})$ pairs, for $i = \{1, \dots, 300\}.$

For the multiple-covariate simulations with non-null association, we generated $z_i$ as $3x_{1i} + 2x_{2i} + u_i$, where either $u_i \sim \mathcal{N}(\mu_{N,q},1)$ or $u_i \sim \mathcal{L}(\mu_{L,q},1)$. As before, we selected values of $\mu_{N,q}$ and $\mu_{L,q}$ so that the $q$th quantile of $\boldsymbol{u}$ would be 0, to give us six different simulations where the ground-truth coefficients are known at a given quantile and error distribution.

For the multiple-covariate simulations with partial null association, we generated $z_i$ as $3x_{1i} + u_i$, where again either $u_i \sim \mathcal{N}(\mu_{N,q},1)$ or $u_i \sim \mathcal{L}(\mu_{L,q},1)$, and we selected $\mu_{N,q}$ and $\mu_{L,q}$ as appropriate. This gives us six different simulations where the ground-truth coefficients are known for each quantile and error distribution, and in particular, the ground-truth coefficient of $x_{2i}$ should be 0 in all cases.

For frequentist continuous quantile regression, we used the \texttt{quantreg} package \cite{koenker2015quantreg} and treated the ordinal response as a continuous outcome. We refer to this method as QR. For AM as well as our modified version of AM, we set the hyperparameters as $\nu = 0.5$, $a_1 = 2.5$, and $a_2 = 4$, following \cite{alhamzawi2017bayesian}. For RM, we set the hyperparameters as $\beta_{p0} = \boldsymbol{0}$, $B_{p0} = \mathbb{I}_p$ (where $\mathbb{I}_p$ denotes the $p \times p$ identity matrix), $c_0 = 5$, and $d_0 = 8$, following suggestions of \cite{rahman2016bayesian}.
For BORPS, we followed the hyperparameter suggestions of \cite{reed2009partially}, and we set $c_0 = d_0 = 10^{-3}$, $\boldsymbol{b_0} = \boldsymbol{0}_p,$ and $B_0 = 10^6 \cdot \mathbb{I}_p,$. 
In all cases, we used 20,000 iterations with a burn-in of 10,000. This was found to be sufficient for convergence and good mixing.
The estimates $\widehat{\boldsymbol{\beta}}$ or $\hat{\frac{\boldsymbol{\beta}}{\delta_{C-1}}}$ as appropriate were determined as the average of the posterior means from 15 runs. For BORPS, to obtain the confidence interval, we used estimates from $100$ bootstrapped samples.

In order to compare results across methods, we computed the root mean-squared error (RMSE) between these point estimates and the ground-truth values. In general, for estimates $\hat{\theta}_i$, $i = 1, \ldots, n$, of a parameter $\theta$, the RMSE is computed as $\sqrt{\frac{1}{n} \sum_{i=1}^n (\hat{\theta}_i - \theta)^2}$. Since we ran each Bayesian method on each simulation $15$ times, the RMSE here was always computed with $n = 15$; for QR, there is only a single point estimate, so $n = 1$. We used $\theta = \beta_k$ or $\theta = \frac{\beta_k}{\delta_{C-1}}$ as appropriate for each method. 

\subsection{Simulation Results}
We evaluated frequentist continuous quantile regression (QR), two Bayesian methods (AM and RM), and our modified version of AM, alongside our method BORPS on the simulated datasets. 

\begin{figure}
\centering
\includegraphics[width=\textwidth]{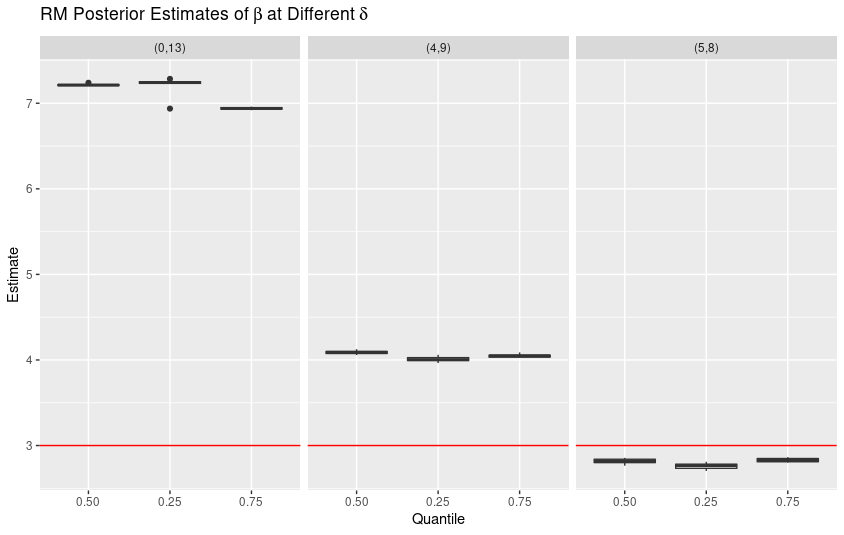}
\caption{Sensitivity of RM to the given cutpoint values ($\delta_1$ and $\delta_2$) in simulated data, when the cutpoint values are correctly specified (right), slightly misspecified (middle), and dramatically misspecified (left).}
\label{fig:rm_ests}
\end{figure}

Unlike the other methods, RM requires the value of the cutpoint vector $\boldsymbol{\delta}$ to be fixed ahead of time, rather than estimated; we assessed the sensitivity of RM's results to the preset $\boldsymbol{\delta}$ value in our first simulation. To do this, we compared the simulated models at each quantile to three cases of RM: when $\boldsymbol{\delta}$ is set to the simulated value ($\delta_1 = 5, \delta_2 = 8$), when it is slightly misspecified ($\delta_1 = 4, \delta_2 = 9$), and when it is dramatically misspecified ($\delta_1 = 0, \delta_2 = 13$). The coefficient estimates are sensitive to this specification of $\boldsymbol{\delta}$ (Figure~\ref{fig:rm_ests}). While the RMSE in the well-specified case is less than $0.20$ for each quantile, it jumps to around one when slightly misspecified, and then to around four when dramatically misspecified.

The sensitivity of RM to the set values of $\boldsymbol{\delta}$ limits its ability to be used in any non-simulated scenario. From the samples alone, it is not clear how to determine appropriate values of $\delta$, especially in a high-dimensional setting. Treating the values of $\boldsymbol{\delta}$ as hyperparameters and holding out a validation set to select appropriate values would also be challenging: This further limits the amount of training data available, and a grid search over appropriate grids would be both computationally challenging and imprecise.

Both AM and BORPS (and by extension, our modified version of AM) avoid the challenge of setting cutpoints by estimating $\boldsymbol{\delta}$ as part of the statistical inference. Because there is no principled way to set $\boldsymbol{\delta}$ as hyperparameters for RM, and because we have demonstrated that the model is sensitive to these values, we focus our attention in the rest of this section on comparing the performances of AM, our modified version of AM, and BORPS, with QR as a baseline. 

\begin{figure}
    \centering
    \includegraphics[width=\textwidth]{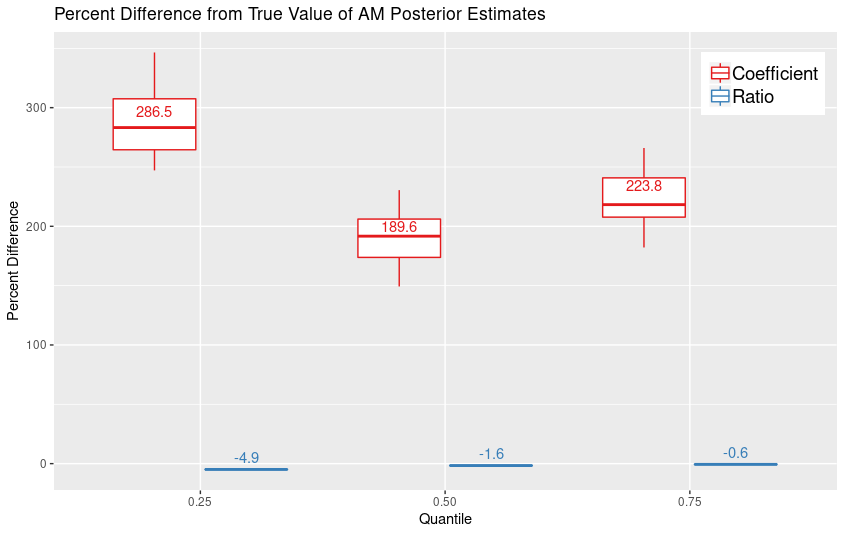}
    \caption{Percent differences between the posterior means from 15 runs of AM and the simulated values, for the absolute coefficients $\beta_1$ (AM) and the ratio $\frac{\beta_1}{\delta_2}$ (our modified version of AM) in a single-covariate simulation with normal errors and non-null association. The average percent difference for each estimated value and quantile is shown above the corresponding box. A percent difference of 0 corresponds to the simulated value.}
    \label{fig:AM_coefs}
\end{figure}

A chief distinction of AM is that this method is intended to report the absolute coefficients $\bsb$, whereas BORPS reports the ratio $\frac{\bsb}{\delta_{C-1}}$ (fixing the intercept to 0), due to their different approaches to enforcing identifiability. Our modified version of AM mirrors BORPS's approach by also reporting the ratio $\frac{\bsb}{\delta_{C-1}}$ and fixing the intercept to 0. When comparing AM's estimates to our modified AM (mAM), it is clear that our modified version of AM performs much better. The estimates of the posterior means of the coefficients at each quantile from AM deviate substantially from the simulated coefficients, whereas the estimates of $\frac{\bsb}{\delta_{C-1}}$ in our mAM were much closer to the corresponding simulated ratios (Figure~\ref{fig:AM_coefs}). 

Notably, the average percent difference from the posterior mean ratios from the simulated ratios in our mAM was within 5\% at each quantile evaluated, whereas the average percent difference for the coefficients in AM could be as high as over 200\%. This finding is what initially motivated our modification of AM, and a similar pattern was seen across the simulations tested. Hence, our mAM performs much better than AM in its original form.  

\begin{figure}
\centering
\includegraphics[width=\textwidth]{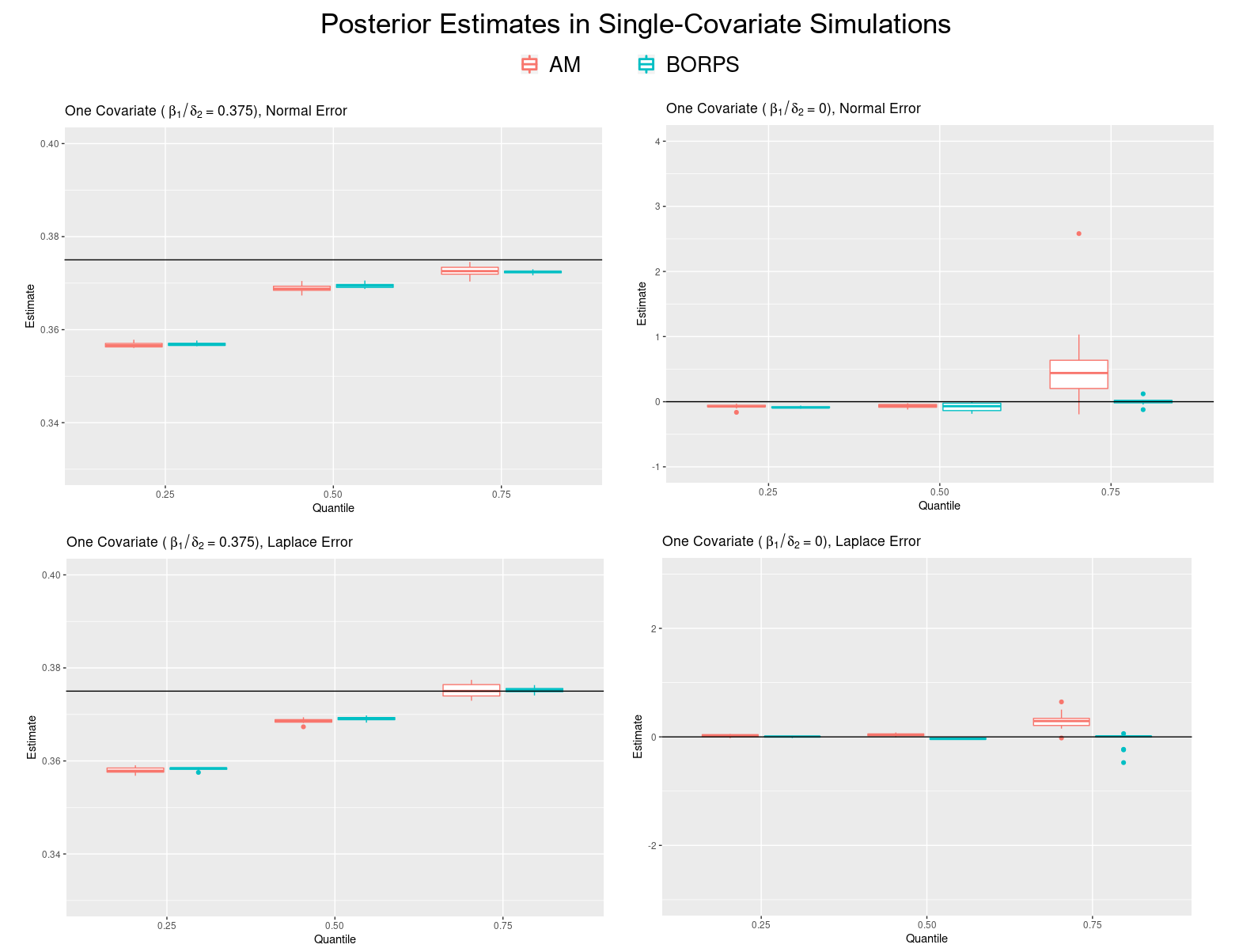}
\caption{Posterior means of $\frac{\beta}{\delta_2}$ over 15 runs from our modified version of AM and BORPS for the single-covariate simulations with two different error distributions (normal, top; Laplace, bottom) and the cases of non-null association (left) and null association (right). Simulated values are indicated by the black horizontal lines in each figure.}
\label{fig:AM_BORPS_singlecov}
\end{figure}

\begin{figure}
\centering
\includegraphics[width=\textwidth]{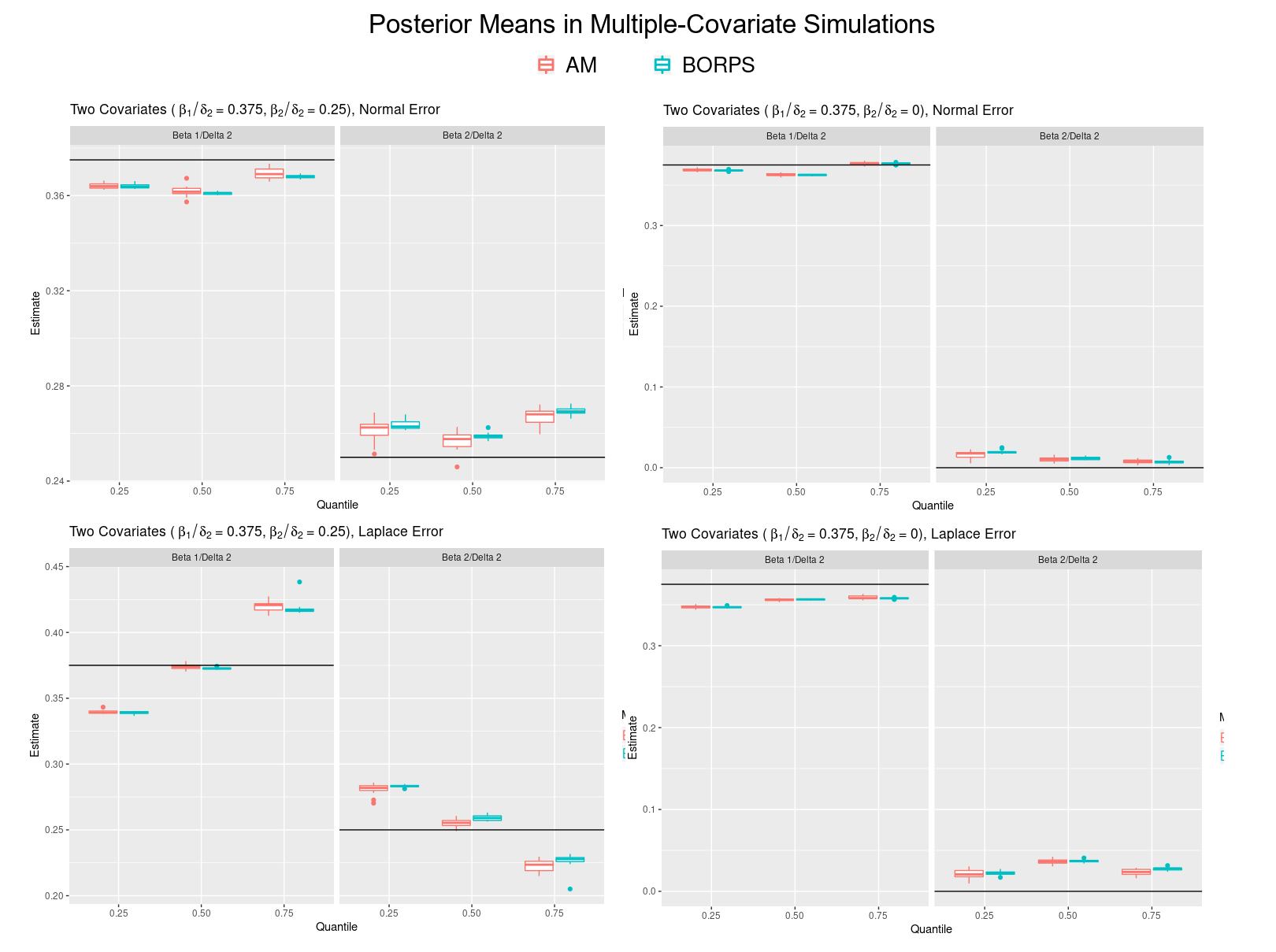}
\caption{Posterior means of $\frac{\bsb}{\delta_2}$ over 15 runs from our modified version of AM and BORPS for the multiple-covariate simulations with two different error distributions (normal, top; Laplace, bottom) and the cases of non-null association (left) and partial null association (right). Simulated values are indicated by the black horizontal lines in each figure.}
\label{fig:AM_BORPS_multicov}
\end{figure}

We have shown already that the two existing methods, AM and RM, both have shortcomings. If we then compare our modified version of AM to BORPS on both single-covariate simulations (Figure~\ref{fig:AM_BORPS_singlecov}) and multiple-covariate simulations (Figure~\ref{fig:AM_BORPS_multicov}), in many cases, the two methods achieve similar RMSEs (Tables 1, 2). However, BORPS is superior to mAM in two important ways. First, mAM sometimes produces especially poor estimates in the case of null association, particularly at extreme quantiles of the single-covariate simulations. By contrast, BORPS continues to reliably and accurately estimate parameters in the null and partially null setting. For example, in the single-covariate case of null association at the 0.75th quantile, our RMSE was less than half the RMSE of mAM for both error distributions. Second, due to the partially collapsed nature of BORPS's sampler, BORPS is able to produce estimates more quickly than mAM. Hence, BORPS is preferable to AM, RM, and mAM.

As a baseline, we also show RMSEs for QR, the frequentist continuous method. QR actually performs well in the null cases, but unsurprisingly yields high RMSEs in all other situations. This makes sense, because QR is handicapped by its assumption that responses are continuous rather than ordinal. This should not affect its performance in the null and partially null cases, which is why it performs well, but it results in incorrect estimates in all other settings. This justifies our use of methods tailored to ordinal data. 

\begin{table}[]
\centering
\begin{tabular}{|l|l|l|l|l|l|l|}
\hline
         & \multicolumn{3}{c|}{\begin{tabular}[c]{@{}c@{}}Normal Error,\\ Non-Null\end{tabular}} & \multicolumn{3}{c|}{\begin{tabular}[c]{@{}c@{}}Laplace Error,\\ Non-Null\end{tabular}} \\ \hline
Quantile & 0.25                        & 0.50                       & 0.75                       & 0.25                        & 0.50                        & 0.75                       \\ \hline \hline 
Modified AM       & 0.0183             & 0.0062                     & 0.0028                     & 0.0170                      & 0.0064                      & 0.0014                     \\ \hline
BORPS    & 0.0181                     & 0.0056            & 0.0026            & 0.0167             & 0.0059             & 0.0006            \\ \hline
\hline
QR & 2.3535 & 2.3786 & 2.3716 & 2.3477 & 2.3818 & 2.3716 \\ \hline
         & \multicolumn{3}{c|}{\begin{tabular}[c]{@{}c@{}}Normal Error,\\ Null\end{tabular}}     & \multicolumn{3}{c|}{\begin{tabular}[c]{@{}c@{}}Laplace Error,\\ Null\end{tabular}}     \\ \hline
Quantile & 0.25                        & 0.50                       & 0.75                       & 0.25                        & 0.50                        & 0.75                       \\ \hline \hline
Modified AM       & 0.0786                      & 0.0724                     & 1.156                   & 0.0313                      & 0.0467                      & 0.3250                     \\ \hline
BORPS    & 0.0869             & 0.0989            & 0.0490            & 0.0111             & 0.0374             & 0.1508         \\ \hline
\hline
QR & 0 & 0 & 0 & 0 & 0 & 0 \\ \hline
\end{tabular}
\label{tab:RMSE_singlecov}
\caption{Root mean square errors (RMSE) between the average posterior mean estimate of $\frac{\beta}{\delta_2}$ and the simulated value $\frac{\beta}{\delta_2}$ for our modified version of AM and BORPS at each quantile of the single-covariate simulations. We also show RMSE between QR's estimate of $\beta$ and the simulated values $\beta$; hence, in the case of QR, we simply consider $\delta_2=1$.}
\end{table}

\begin{table}[]
\centering
\begin{tabular}{|l|l|l|l|l|l|l|}
\hline
            & \multicolumn{6}{c|}{\begin{tabular}[c]{@{}c@{}}Normal Error,\\ Non-Null\end{tabular}}                     \\ \hline
Coefficient & \multicolumn{3}{c|}{Beta 1/Delta 2}                 & \multicolumn{3}{c|}{Beta 2/Delta 2}                 \\ \hline
Quantile    & 0.25            & 0.50            & 0.75            & 0.25            & 0.50            & 0.75            \\ \hline \hline
Modified AM          & 0.0110 & 0.0134 & 0.0062          & 0.0122          & 0.0080 & 0.0175          \\ \hline
BORPS       & 0.0111          & 0.0140       & 0.0071 & 0.0140 & 0.0090          & 0.0195 \\ \hline
\hline 
QR & 2.3983 & 2.3803 & 2.3666 & 1.7981 & 1.7474 & 1.7014 \\ \hline
            & \multicolumn{6}{c|}{\begin{tabular}[c]{@{}c@{}}Normal Error,\\ Partial Null\end{tabular}}                 \\ \hline
Coefficient & \multicolumn{3}{c|}{Beta 1/Delta 2}                 & \multicolumn{3}{c|}{Beta 2/Delta 2}                 \\ \hline
Quantile    & 0.25            & 0.50            & 0.75            & 0.25            & 0.50            & 0.75            \\ \hline \hline
Modified AM          & 0.0064 & 0.0122         & 0.0024          & 0.0168          & 0.0111          & 0.0082          \\ \hline
BORPS       & 0.0069          & 0.0125 & 0.0021 & 0.0197 & 0.0117 & 0.0076 \\ \hline
\hline 
QR & 2.3316 & 2.3609 & 2.3611 & 0.0281 & 0.0547 & 0.0606 \\ \hline
            & \multicolumn{6}{c|}{\begin{tabular}[c]{@{}c@{}}Laplace Error,\\ Non-Null\end{tabular}}                    \\ \hline
Coefficient & \multicolumn{3}{c|}{Beta 1/Delta 2}                 & \multicolumn{3}{c|}{Beta 2/Delta2}                  \\ \hline
Quantile    & 0.25            & 0.50            & 0.75            & 0.25            & 0.50            & 0.75            \\ \hline \hline
Modified AM          & 0.0354 & 0.0023 & 0.0450          & 0.0312          & 0.0062 & 0.0276          \\ \hline
BORPS       & 0.0360          & 0.0023          & 0.0436 & 0.0332 & 0.0093          & 0.0244 \\ \hline
\hline 
QR & 2.4330 & 2.3825 & 2.3664 & 1.7726 & 1.7620 & 1.7447 \\ \hline
            & \multicolumn{6}{c|}{\begin{tabular}[c]{@{}c@{}}Laplace Error,\\ Partial  Null\end{tabular}}               \\ \hline
Coefficient & \multicolumn{3}{c|}{Beta 1/Delta 2}                 & \multicolumn{3}{c|}{Beta 2/Delta 2}                 \\ \hline
Quantile    & 0.25            & 0.50            & 0.75            & 0.25            & 0.50            & 0.75            \\ \hline \hline
Modified AM          & 0.0278 & 0.0190 & 0.0160          & 0.0219          & 0.0364 & 0.0236          \\ \hline
BORPS       & 0.0279          &0.0186        & 0.0171 & 0.0221 & 0.0372          & 0.0275 \\ \hline
\hline 
QR & 2.3613 & 2.3765 & 2.3893 & 0.0220 & 0.0716 & 0.0484 \\ \hline
\end{tabular}
\label{tab:RMSE_multicov}
\caption{Root mean square errors (RMSE) between the average posterior mean estimates of $\frac{\beta_1}{\delta_2}$ and $\frac{\beta_2}{\delta_2}$ and the corresponding simulated values for AM and BORPS at each quantile of the multiple-covariate simulations. We also show RMSE between QR's estimate of $\beta_1, \beta_2$ and the simulated values $\beta_1, \beta_2$; hence, in the case of QR, we simply consider $\delta_2=1$.}
\end{table}

\begin{figure}
    \centering
    \includegraphics[width=\textwidth]{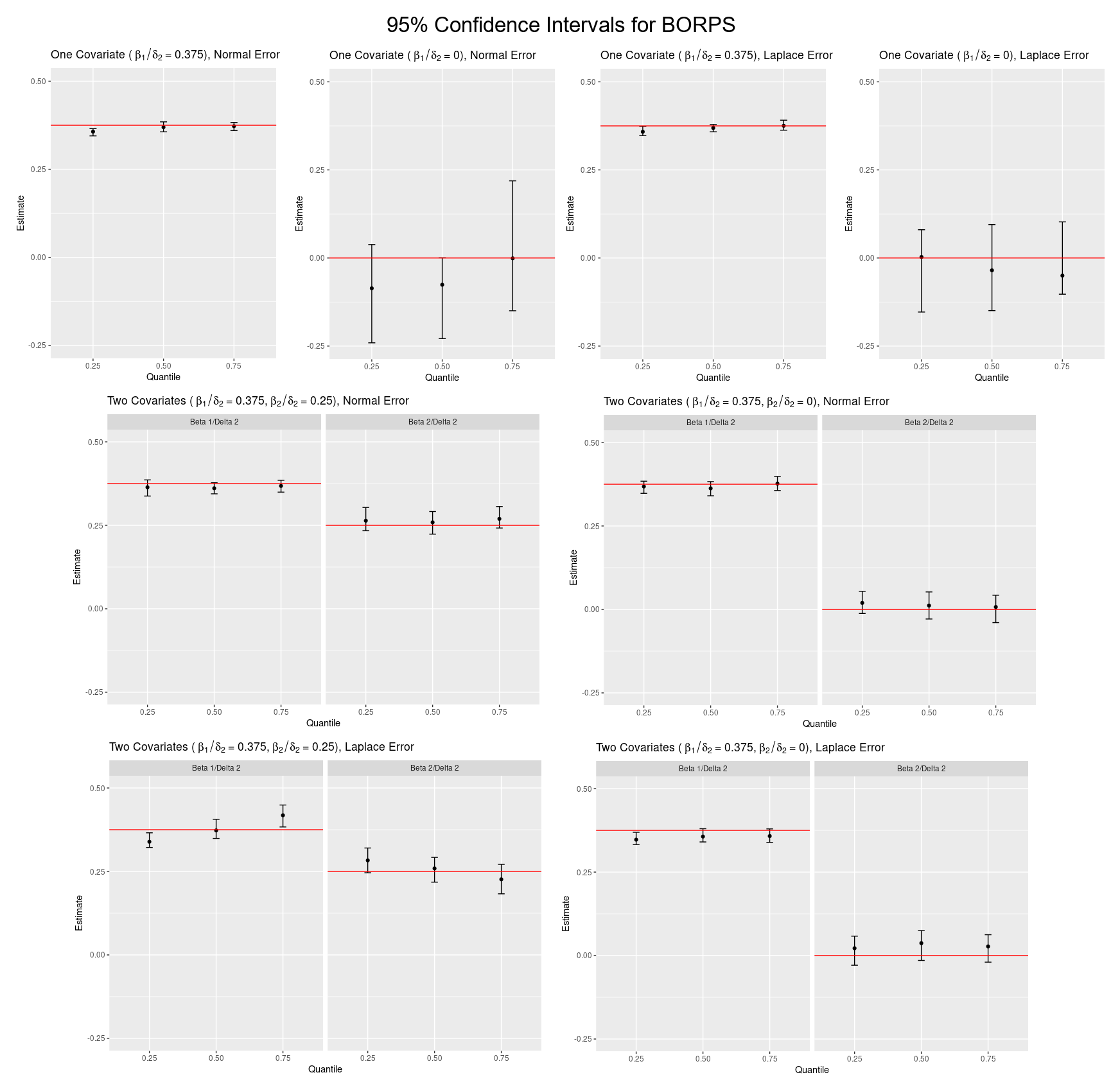}
    \caption{95\% confidence intervals around the posterior means of BORPS estimates, constructed using bootstrapping for each simulation. The red line in each case shows the true values.}
    \label{fig:BORPS_ci}
\end{figure}

Thus far, we have seen that BORPS produces point estimates that are closer to the ground-truth values. Next, we evaluated BORPS's ability to capture these true values in its confidence intervals (CIs) under both the single-covariate and multiple-covariate settings. We constructed 95\% CIs with bootstrapping as described in the previous section, and found that, in the vast majority of cases, these intervals contained the simulated value (Figure~\ref{fig:BORPS_ci}). 

Although there were a few instances where the CI for the 0.25th quantile came close but did not cover the simulated value, it is important to note that the confidence intervals correctly contained or did not contain zero as appropriate across all quantiles and simulated settings. That is, BORPS's CIs include 0 exactly when the simulation response values were generated independently of the covariate in question. This means we learn whether or not $\hat{{\frac{\beta_k}{\delta_{C-1}}}} = 0$ within the CI, which tells us whether or not $\hat{\beta_k}$ differs significantly from $0$, and therefore whether or not we may consider the association statistically significant. 

Taken together, these simulation results suggest that BORPS is capable of producing reliable posterior mean and uncertainty estimates under a variety of situations across quantiles, including in the presence of multiple covariates, noisy data, and zero-valued true coefficients. Hence, BORPS's functionality is two-fold. First, we can compare the relative associations between the response variable and its covariates across quantiles; we have demonstrated more accurate estimates than existing methods, especially at more extreme (non-median) quantiles. Second, we can evaluate whether the association with a given covariate at a given quantile is significant or not; by contrast, the existing methods we examined do not give any procedure to determine the statistical significance of their associations. In particular, AM demonstrated poor coefficient estimates in some data simulated under a null association.

\section{Application to Early Puberty}
As an example application of our method, we use BORPS to study early puberty from the Fragile Families Childhood Wellbeing Study (FFCWS), a longitudinal study following nearly 5,000 children and their families over fifteen years with an intentional overrepresentation of low-income and minority children \cite{reichman2001fragile}. Families were interviewed at the time of the focus child's birth and again when the child was age 1, 3, 5, 9, and 15. The households of FFCWS participants include heterogeneous family dynamics and conditions, which enables the study of many phenotypes of interest from a range of complex childhood environments. 

The average age of female puberty onset in the United States is trending downwards, particularly in low-income and African-American populations \cite{krieger2015age}, but the underlying reasons are not well understood. Previous research has identified links between early puberty and socioeconomic factors \cite{jansen2015trends,deardorff2014socioeconomic}, harsh family environments \cite{sung2016secure}, nutrition \cite{carwile2015sugar,jansen2015higher}, and childhood obesity \cite{yermachenko2014nongenetic,al2013age}, but there are many contradicting studies, and no clear consensus in the literature. 

These links are important to understand because early puberty is associated with increased health risks later in life, including for cardiovascular disease, breast cancer, and type 2 diabetes \cite{perry2014parent}. In fact, the relative risk of all-cause mortality increases with earlier pubertal onset \cite{charalampopoulos2014age}. The mechanisms linking early puberty and these increased health risks later in life are no better understood than the causes of early puberty itself, so identifying childhood factors associated with early puberty is essential to hypothesize causal models that lead to developing effective intervention strategies. 

Much of the research on early puberty has focused on girls, with less investigation on the associations and consequences of early puberty in boys. Hence, it is also of interest as well to see if any childhood factors are related to early puberty in boys.

Quantile regression offers a new perspective for this problem, by allowing the conditional quantiles---not only the conditional mean---to be modeled. Two features of these data make quantile regression a compelling modeling approach. First, the relationships among the various potential factors are complicated, change based on quantile, and are highly non-Gaussian. Second, the survey data, like many large-scale surveys, contain errors, incorrect reporting, and sample outliers. Thus, quantile regression is useful as a way to both explore changing associations across quantiles and to mediate the effects of the prevalent misclassification and measurement error in the data. Like many health outcomes, the puberty measures in these data are ordinal variables, motivating the use of BORPS. 

We examine year 9 of the FFCWS data and use the \emph{breast development} variable to represent early puberty in girls, and the \emph{growth of underarm/pubic hair} variable to represent early puberty in boys. Breast development onset is considered one way to determine the beginning of puberty in girls \cite{mouritsen2013pubertal}, and such onset at age 9 is typically considered early \cite{aksglaede2009recent}. Although the beginning signs of puberty tend to be less overt in boys than in girls, the onset of pubic hair at age 9 would be considered early in boys \cite{pinyerd2005puberty}, motivating our use of this variable. 

In the FFCWS data, breast development and growth of underarm/pubic hair are both coded ordinally from 1 to 4, where the lowest value of 1 indicates ``no development,'' a value of 2 indicates ``barely developed,'' a value of 3 indicates ``definitely developed,'' and the highest value of 4 indicates ``that development is already completed.'' In both cases, there were so few responses with a value of 4 that we only included responses ranging from 1 to 3 inclusive. 

\begin{figure}
    \includegraphics[width=0.46 \textwidth, trim={2cm 0 0 0}]{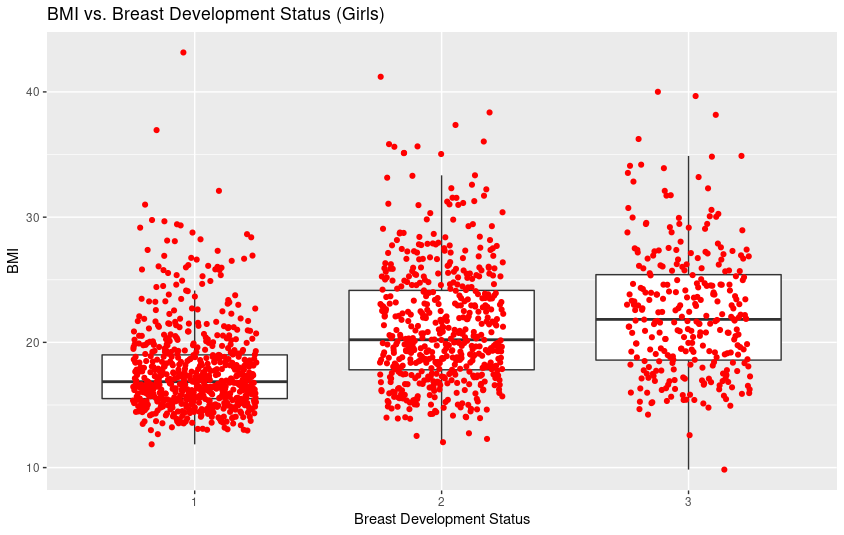}
    \includegraphics[width=0.5\textwidth]{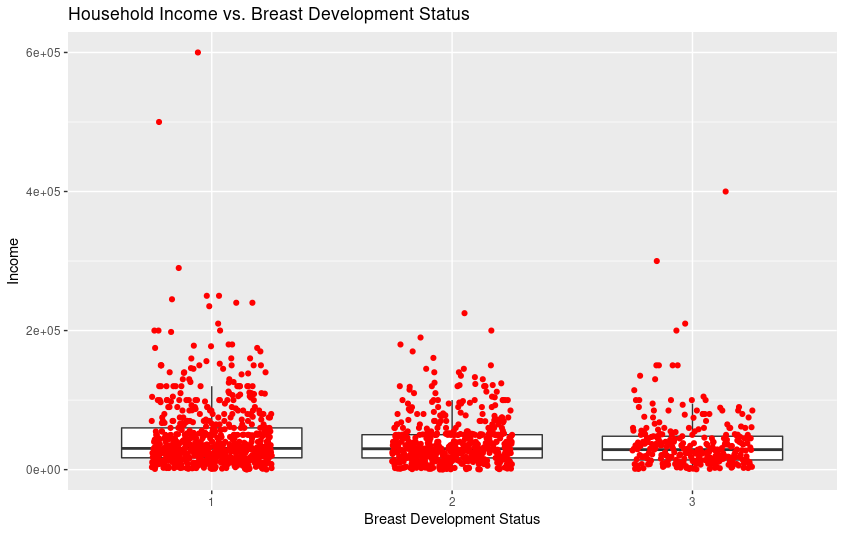}\\
   \includegraphics[width=0.5\textwidth]{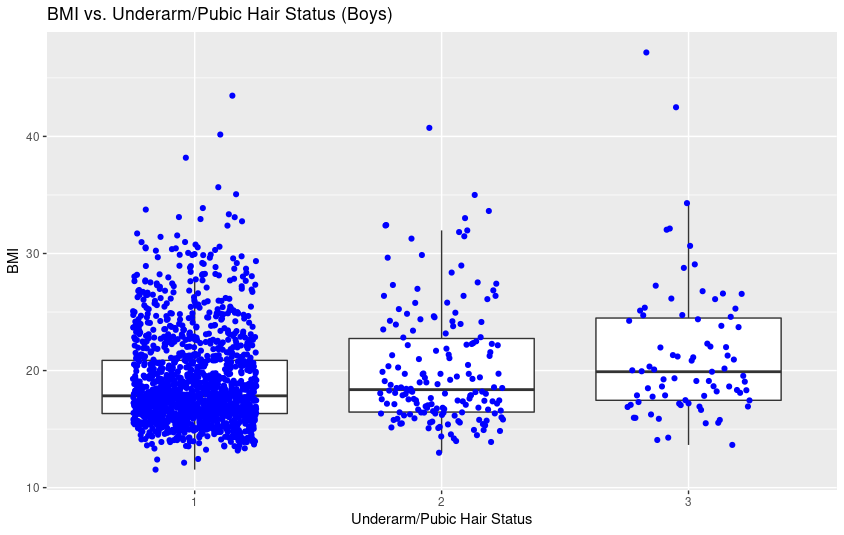}
    \includegraphics[width=0.5\textwidth]{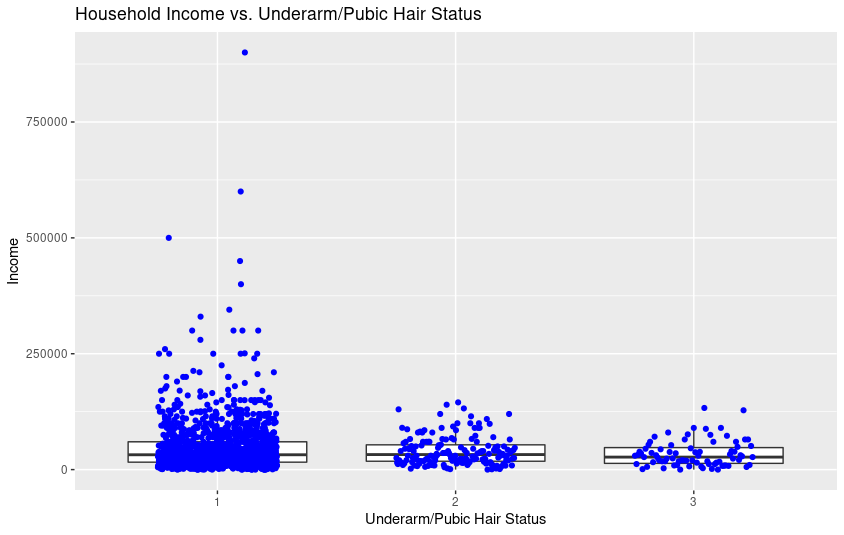}
    \caption{BMI and household income for FFCWS participants at each of the three possible early puberty statuses, corresponding to ``no development'', ``somewhat developed'', and ``definitely developed''. The top figures (in red) refer to breast development in girls, and the bottom figures (in blue) refer to underarm/pubic hair in boys.}
    \label{fig:pub_boxplots}
\end{figure}

We choose two covariates to consider as potentially associated with breast development (in girls) and growth of underarm/pubic hair (boys) at age 9, based on the conflicting evidence from previous research described above: age-adjusted BMI and household income (Figure~\ref{fig:pub_boxplots}). Because of the differing scale of the covariates, we choose to normalize them for our analysis here as z-scores; this is standard in regular (non-quantile) ordinal regression and is recommended, for example, by the popular package \texttt{MASS} \cite{MASS}.

As a baseline, we applied two conventional methods to study this data: the ordinal probit model with the R package \texttt{MASS} \cite{MASS}, and frequentist quantile regression (QR) with the R package \texttt{quantreg} \cite{koenker2015quantreg}. The ordinal probit model uses a similar latent variable framework as BORPS to model the ordinal responses, but assumes a standard normal distribution for the residuals and estimates coefficients only at the mean of the conditional response distribution. QR has been discussed in earlier sections; unlike the ordinal probit model, it carries out inference at any quantile of interest in the conditional response distribution, but can only treat responses as continuous values. We applied it with its default settings.

When the ordinal probit model was applied to breast development in girls, this model yielded a coefficient of approximately $0.48$ for BMI and $-0.07$ for household income, with significant associations in both cases ($p < 0.05$ for both covariates). The sign of both effects are consistent with our understanding of these relationships: higher BMI appears to be associated with more substantial breast development (positive effects), and lower income appears to be associated with more substantial breast development (negative effects). 

For underarm/pubic hair development in boys, the ordinal probit model yielded a coefficient of approximately $0.17$ for BMI and $-0.12$ for household income, with significant associations again in both cases ($p < 0.05$). The signs of these effects are the same as for breast development in girls, albeit with a smaller effect size for BMI.

We applied QR at five quantiles ($0.05, 0.25, 0.50, 0.75$, and $0.95$). For breast development in girls, at the $0.50$th quantile, the estimates were similar to the findings of the ordinal probit model, with a coefficient of approximately $0.45$ for BMI and $-0.04$ for household income (both $p < 0.05$). The coefficients were similar at the 0.25th and 0.75th quantile, without a clear trend across these middle three quantiles, except for the one notable difference that household income was no longer significant at the 0.25th quantile. Interestingly, at the two extreme quantiles (0.05 and 0.95), the estimated coefficients were reported as exactly 0 for both BMI and income, and standard errors could not be computed. Hence, at these extremes, QR was unable to carry out meaningful inference.

For underarm/pubic hair development in boys, QR found a general trend of the coefficients for both BMI and income to become more negative with increasing quantile. At the 0.50th quantile, the coefficient for BMI was approximately $-0.027$, and the coefficient for income was approximately $-0.034$. This is different from what was found by the ordinal probit model, most notably because BMI had a negative association. The effect of income was significant at each quantile evaluated greater than the 0.50th quantile, and BMI only had a significant effect at the 0.95th quantile. Unlike for breast development in girls, there were no issues with inference at the extremes. 

Next, we applied BORPS to see if there was information to be gained from estimating the conditional quantiles under our approach (Figure~\ref{fig:bmi-income}). For breast development in girls, the values found by BORPS at the median were similar to what QR had found at the median, as well as to the coefficients estimated by the ordinal probit model. Namely, the posterior mean estimates for BORPS was approximately $0.48$ for BMI and $-0.08$ for household income, as compared to $0.45$ and $-0.04$ respectively from QR, and $0.48$ and $-0.07$ respectively from the ordinal probit model. Further, like these two models, BORPS found both of these associations to be significant (with neither confidence interval overlapping 0).

\begin{figure}
\centering
\includegraphics[width=0.49\textwidth]{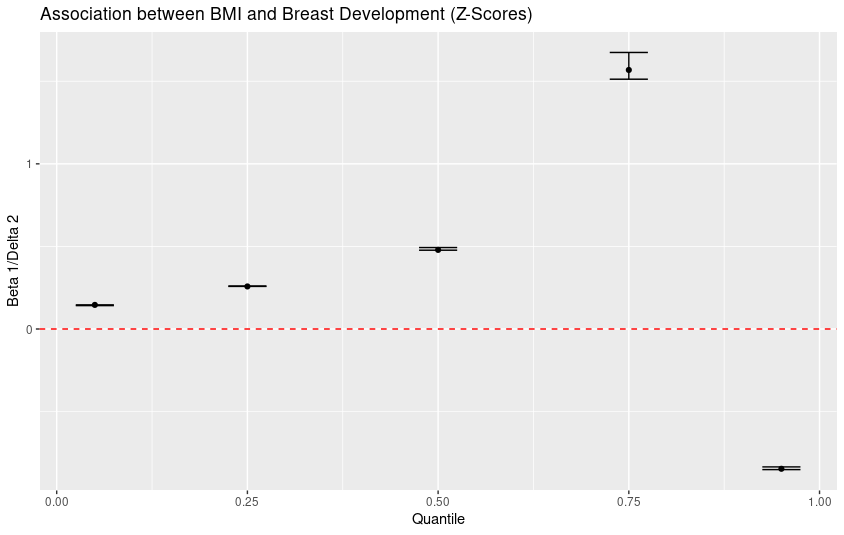}
\includegraphics[width=0.49\textwidth]{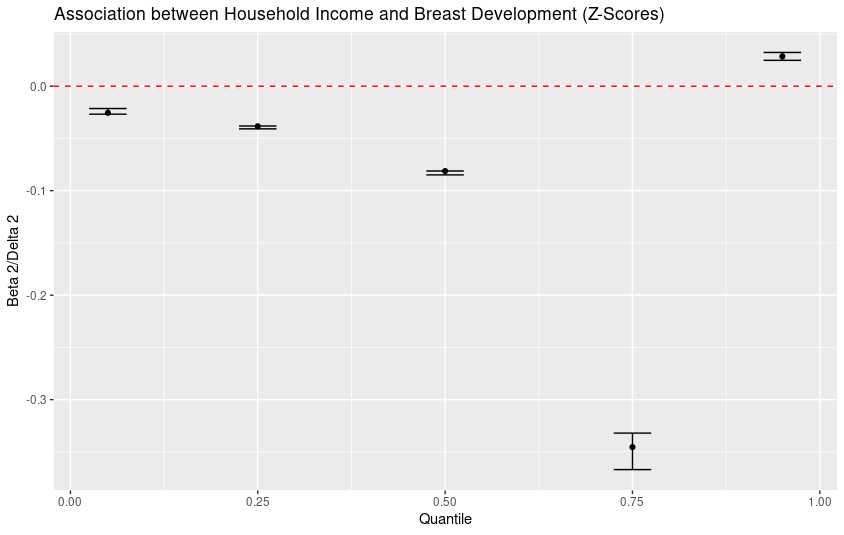}
\caption{Coefficients found by BORPS across quantiles for BMI (left) and income (right) for breast development in girls, using z-scored covariates. The x-axis represents five quantiles, and the y-axis represents the ratio $\frac{\beta}{\delta_2}$ for the appropriate coefficient, with the points representing posterior mean estimates and whiskers representing 95\% confidence intervals. The red dashed line represents the 0 line.} 
\label{fig:bmi-income}
\end{figure}

\begin{figure}
\centering
\includegraphics[width=0.49\textwidth]{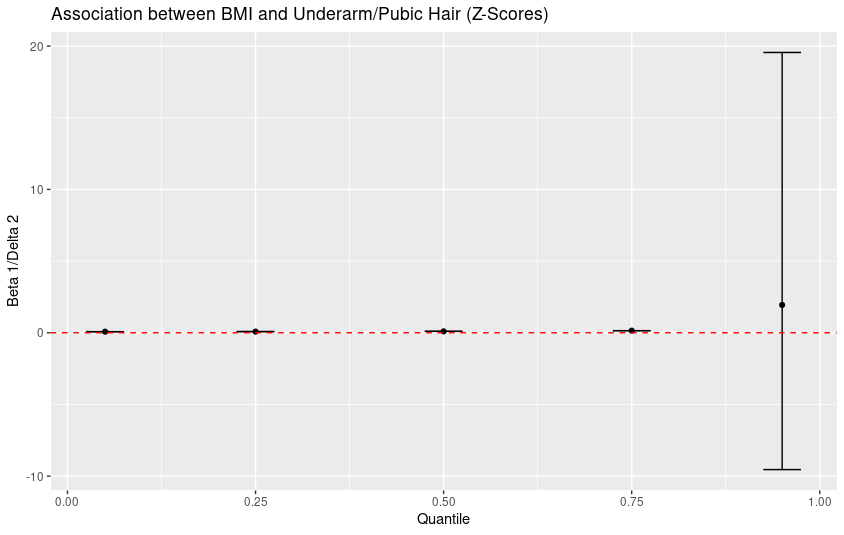}
\includegraphics[width=0.49\textwidth]{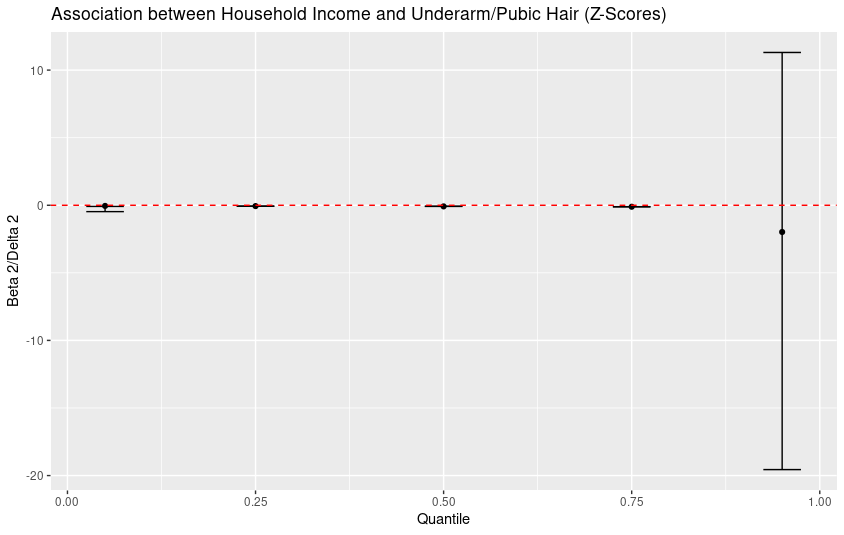} 
\caption{Coefficients found by BORPS across quantiles for BMI (left) and income (right) for underarm/pubic hair in boys, using z-scored covariates. The x-axis represents five quantiles, and the y-axis represents the ratio $\frac{\beta}{\delta_2}$ for the appropriate coefficient, with the points representing posterior mean estimates and whiskers representing 95\% confidence intervals. The red dashed line represents the 0 line.} 
\label{fig:bmi-income-boys}
\end{figure}

In contrast to what was observed with QR, BORPS's estimated effect sizes follow a trend over the quantiles for both covariates (Figure~\ref{fig:bmi-income}). As the quantile increases, the estimate of effect size for BMI grows up the 0.75th quantile, at which point it changes sign. This suggests that at higher quantiles of the conditional breast development distribution, BMI has a greater positive effect on breast development, until the most extreme quantiles (i.e., the 0.95th quantile), at which point BMI has a significant negative effect. This result might lead us to speculate about a biological mechanism in which BMI plays an increasingly important role as premature breast development becomes more pronounced. However, as indicated by our result at the 0.95th quantile, an entirely different biological mechanism might apply to the extreme cases of premature breast development, in which BMI instead negatively affects development. 

A similar pattern but in reverse is seen for income. The association between income and early puberty becomes increasingly negative as the quantile increases until the 0.75th quantile, at which point it becomes a slight positive association. This has an analogous interpretation as above, suggesting that household income has a negative effect on early breast development, except at the most extreme quantile (0.95th) considered. This finding is consistent with our speculation of a distinct mechanism governing extreme cases of early breast development. For all but the most extreme cases, a mechanism may be at play that causes more breast development as BMI increases and household income decreases, but then the opposite might hold true at the upper quantiles.

For underarm/pubic hair in boys, the coefficients at the median were again similar to the coefficients found by ordinal probit model, at $\frac{\beta_1}{\delta_2} = 0.10$ for BMI (compared to $0.17$ from the ordinal probit model) and $\frac{\beta_2}{\delta_2} = -0.08$ for household income (compared to $-0.12$ from the ordinal probit model). The fact that there is still some difference between BORPS's values and the ordinal probit model suggests that the conditional response distribution is not symmetric, which is why our estimate at the median would differ from an estimate at the mean. Nevertheless, there was much closer agreement between our results and the ordinal probit model than there was between QR and the ordinal probit model. This again supports the idea that treating these outcomes as ordinal, rather than continuous, is important for inference.

Over the rest of the quantiles, the values for BMI increased with increasing quantile, whereas the values of household income remained roughly the same (Figure~\ref{fig:bmi-income-boys}). This suggests that, at higher quantiles of the conditional underarm/pubic hair distribution, BMI has an increasing effect, whereas household income has a steady, negative effect. The exception is at the 0.95th quantile, which is the only quantile for both covariates where there is a non-significant effect (i.e., the confidence interval overlaps 0). It may be possible, then, that the most extreme cases of early underarm/pubic hair development may be attributed to some other biological mechanism that is not associated with BMI and household income. 

We also performed this analysis with unnormalized covariates (see Section 2 of the supplement), even though this is not recommended in ordinal regression. In this setting, we achieved a similar trend of results, except, in all cases, the coefficient changed sign at the 0.95th quantile from what was observed when we used normalized covariates. This may suggest that BORPS has some sensitivity to scale at very extreme quantiles, or this may also simply be reflective of the very little information available at such extremes. QR, for instance, was entirely unable to perform inference at this extreme quantile for breast development in girls, since standard errors could not even be computed. 

Overall, BORPS has given us a wider range of insight into this problem than the narrow slice of information from the ordinal probit model. Moreover, results from BORPS agree, when applicable, with results from the ordinal probit model, but this is not always the case for QR, which supports the need for modeling ordinal outcomes appropriately. BORPS allows us to capture meaningful but complex relationships among covariates in the tails of these conditional distributions, overcoming the limitations of models that only examine the conditional mean or that treat these outcomes as continuous.

\section*{Discussion}

Motivated by the study of associations of specific covariates with early puberty as an ordinal response in the Fragile Families Study, we developed a method for Bayesian ordinal quantile regression using a partially collapsed Gibbs sampler (BORPS). BORPS improves on existing methods by both robustly estimating quantile regression parameters and allowing full Bayesian inference using a fast collapsed Gibbs sampler. The resulting methods leads to statistical tests of association with the ordinal response and the model covariates. 

We validated BORPS against two state-of-the-art quantile regression methods on simulated data to show precise and robust modeling of the data relative to the other methods. Next, we used BORPS to model early puberty associations in the Fragile Families cohort. We found that BMI and income are both associated with early puberty, with different directionality at the extreme upper tail as from the remaining quantiles. This suggests a distinct biological mechanism governing extreme early breast development as from the rest of the population. We have released BORPS as open source software for use in ordinal quantile regression.

\section*{Acknowledgments}

 BEE was supported by NIH R01 HL133218 and an NSF CAREER AWD1005627. ING was supported by the National Library Of Medicine of the National Institutes of Health under Award Number T32LM012411. The content is solely the responsibility of the authors and does not necessarily represent the official views of the National Institutes of Health. We gratefully acknowledge the Fragile Families Childhood Wellbeing Study participants and scientists, and Profs Sara McLanahan and Daniel Notterman in particular, for their important role in this work.
 
{\it Conflict of Interest}: BEE is on the SAB for Celsius Therapeutics and Freenome, and is currently employed by Genomics plc and Freenome and on a year leave-of-absence from Princeton University.

\bibliographystyle{unsrt}
\bibliography{samplebibtex.bbl}

\section{Supplementary: Gibbs Sampler Derivations for BORPS}
As described in the Methods section, we denote our observed response variables $y_i$, for $n$ samples indexed by $i \in \{1,\dots,n\}$, with $p$ corresponding covariates $\boldsymbol{x_i}$. Each $y_i$ is related to a continuous latent $z_i$ via the cutpoint vector $\boldsymbol{\delta}$, and we model $z_i = \boldsymbol{\beta}^T \boldsymbol{x_i} + \epsilon_i$ for coefficient $\boldsymbol{\beta} = \big[ \beta_1, \ldots, \beta_k \big]$ and residual $\epsilon_i \sim \mathcal{ALD}(0,\sigma,q)$, where $0 < q < 1$ is our quantile of interest. 

Our model can be rewritten as $z_i = x_i'\beta + \sigma \theta w_i + \sigma \tau \sqrt{w_i} u_i,$ using the normal-exponential mixture of the $\mathcal{ALD}$ \cite{rahman2016bayesian}. Here, $w_i \sim \mathcal{E}(1)$, $u_i \sim \mathcal{N}(0,1)$, $\theta = \frac{1-2q}{q(1-q)}$, and $\tau = \sqrt{\frac{2}{q(1-q)}}$. This means $z_i \sim \mathcal{N}(x_i'\beta + \theta v_i, \tau^2 \sigma v_i)$ for $v_i = \sigma w_i$. As in prior work \cite{rahman2016bayesian,reed2009partially}, we place a Gamma prior on $\sigma^{-1}| c_0, d_0 \sim \Gamma(c_0,d_0)$, a Gamma prior on $\nu_i | \sigma \sim \Gamma(1,\sigma^{-1})$, and a multivariate normal prior on $\beta | \boldsymbol{b_0},B_0 \sim \mathcal{MVN}(\boldsymbol{b_0},B_0)$. We also use an order statistic from the uniform distribution for $\boldsymbol{\delta}$ \cite{alhamzawi2017bayesian}, i.e., $P(\delta) = (C-1)! \left( \frac{1}{\delta_C - \delta_0} \right) \mathbb{I}(\delta \in T)$ when $T = \{ (\delta_0, \ldots, \delta_C) | \delta_0 < \cdots < \delta_C \}$. 

To construct the sampler, we nearly directly follow the derivations for the continuous counterpart \cite{reed2009partially}; we walk through their derivations below, but with our notation. First, we define $u_i = z_i - \theta v_i$, and $\boldsymbol{V}$ has diagonal elements $v_i^{-1}$. Then the full likelihood is \begin{equation*} l(\boldsymbol{z}|\boldsymbol{\beta}, \sigma, \boldsymbol{V}) \propto \sigma^{-\frac{n}{2}} \left(\prod_{i=1}^n v_i^{-\frac{1}{2}}\right) \exp \left(-\frac{(\boldsymbol{u}-\boldsymbol{X}\boldsymbol{\beta})^T \boldsymbol{V} (\boldsymbol{u}-\boldsymbol{X}\boldsymbol{\beta})}{2\tau^2 \sigma}\right), \end{equation*} so the posterior distribution is  \begin{align*} \pi(\boldsymbol{\beta},\sigma,\boldsymbol{V}|\boldsymbol{z}) &\propto l(\boldsymbol{z}|\boldsymbol{\beta}, \sigma, \boldsymbol{V}) \pi(\boldsymbol{V}|\boldsymbol{\beta},\sigma) \pi(\boldsymbol{\beta}|\sigma) \pi(\sigma) \\ &\propto \sigma^{-\frac{n}{2}} \left(\prod_{i=1}^n v_i^{-\frac{1}{2}}\right) \exp \left(-\frac{(\boldsymbol{u}-\boldsymbol{X}\boldsymbol{\beta})^T \boldsymbol{V} (\boldsymbol{u}-\boldsymbol{X}\boldsymbol{\beta})}{2\tau^2 \sigma}\right) \\ &\times \left(\prod_{i=1}^n v_i^2\right) \sigma^{-n} \exp\left(-\frac{1}{\sigma}\sum_{i=1}^n v_i\right) \\ &\times \exp\left(-\frac{1}{2}(\boldsymbol{\beta}-\boldsymbol{b}_0)^TB_0(\boldsymbol{\beta}-\boldsymbol{b}_0)\right) \times \sigma^{1-c_0} \exp\left(-\frac{d_0}{\sigma}\right). \end{align*}

To make this a partially collapsed sampler, we integrate out $\boldsymbol{V}$ to reduce the conditional posterior for $\sigma^{-1}$. This is done by multiplying our prior for $\sigma^{-1}$ by the reduced likelihood \begin{equation*} l(\boldsymbol{z}|\boldsymbol{\beta},\sigma) \propto \sigma^{-n} \exp\left(-\frac{\sum_{i=1}^n \rho_q(z_i - \boldsymbol{x}_i^T\boldsymbol{\beta}}{\sigma}\right), \end{equation*} which comes from directly using the $\mathcal{ALD}$ likelihood instead of the normal-exponential mixture. This yields \begin{equation*} \pi(\sigma^{-1}|\boldsymbol{\beta},\boldsymbol{z}) \propto \sigma^{-n} \exp\left(-\frac{\sum_{i=1}^n \rho_q(z_i - \boldsymbol{x}_i^T\boldsymbol{\beta}}{\sigma}\right) \times \sigma^{1-c_0} \exp\left(-\frac{d_0}{\sigma}\right). \end{equation*} We can then say that \begin{equation*} \sigma^{-1} | \boldsymbol{\beta},\boldsymbol{z} \sim \Gamma\left(c_0 + n, d_0 + \sum_{i=1}^n \rho_q(z_i - \boldsymbol{x}_i^T \boldsymbol{\beta})\right). \end{equation*}

For $\boldsymbol{\beta},$ we can complete the square to find \begin{equation*} \boldsymbol{\beta} | \sigma^{-1}, \boldsymbol{V}, \boldsymbol{z} \sim \mathcal{N}\left(\hat{\boldsymbol{\beta}}, \left(\frac{1}{\tau^2 \sigma}\boldsymbol{X}^T\boldsymbol{V}\boldsymbol{X} + \boldsymbol{B}_0^{-1}\right)^{-1}\right) \end{equation*} for \begin{equation*} \hat{\boldsymbol{\beta}} = \left(\frac{1}{\tau^2 \sigma} \boldsymbol{X}^T \boldsymbol{V}\boldsymbol{X} + \boldsymbol{B}_0^{-1}\right)^{-1} \left(\frac{1}{\tau^2 \sigma}\boldsymbol{X}^T\boldsymbol{V}\boldsymbol{u} + \boldsymbol{B}_0^{-1} \boldsymbol{b}_0\right). \end{equation*}

For $\boldsymbol{V}$, we can find the conditional posterior for each $v_i^{-1}$ as  \begin{align*} \pi(v_i^{-1} | \boldsymbol{\beta},\sigma,\boldsymbol{z}) &\propto v_i^{\frac{3}{2}} \exp\left(-\frac{1}{2\tau^2 \sigma v_i} \left(z_i - \boldsymbol{x}_i^T \boldsymbol{\beta} - \theta v_i\right)^2 + \frac{v_i}{\sigma}\right) \\ &\propto v_i^{\frac{3}{2}}\exp\left(-\frac{(z_i - \boldsymbol{x}_i^T \boldsymbol{\beta})^2}{2\tau^2 \sigma v_i} - v_i\left(\frac{\theta^2}{2\tau^2 \sigma} + \frac{1}{\sigma}\right)\right) \\ &= v_i^{-\frac{3}{2}} \exp\left(-\frac{(z_i - \boldsymbol{x}_i^T \boldsymbol{\beta})^2}{2\tau^2 \sigma v_i} - \frac{v_i \tau^2}{8\sigma}\right), \end{align*} which means that each \begin{equation*} v_i^{-1} | \boldsymbol{\beta}, \sigma, \boldsymbol{z} \sim \mathcal{IG}\left(\frac{1}{q(1-q)|z_i - \boldsymbol{x}_i^T \boldsymbol{\beta}|}, \frac{\tau^2}{4 \sigma}\right), \end{equation*} where $\mathcal{IG}$ denotes the inverse Gaussian distribution.

For $\boldsymbol{z}$, we follow the work on AM and RM \cite{alhamzawi2017bayesian,rahman2016bayesian} to write \begin{align*} \pi(z_i|\boldsymbol{\beta},\boldsymbol{\delta},v_i,\sigma) &\propto l(y_i|z_i,\boldsymbol{\delta}) \pi(z_i|\boldsymbol{\beta},\sigma,v_i) \\ &\propto \mathbb{I}(\delta_{c-1} < z_i \leq \delta_c)\mathcal{N}(\boldsymbol{x_i}^T\boldsymbol{\beta}+ \theta v_i, \tau^2 \sigma v_i). \end{align*} This means that \begin{equation*} z_i \sim \mathcal{TN}_(\delta_{c-1},\delta_c)(\boldsymbol{x_i}^T\boldsymbol{\beta}+ \theta v_i, \tau^2 \sigma v_i), \end{equation*} where $\mathcal{TN}$ denotes a truncated normal distribution.

For $\boldsymbol{\delta}$, we again follow the work on AM \cite{alhamzawi2017bayesian} to say that each \begin{equation*} \pi(\delta_c | \boldsymbol{y}, \boldsymbol{z}) \propto \prod_{i=1}^n \sum_{c=1}^C \mathbb{I}(y_i = c) \mathbb{I}(\delta_{c-1} < z_i < \delta_c) \mathbb{I}(\boldsymbol{\delta} \in T), \end{equation*} where $T$ is is the set of $\boldsymbol{\delta}$ satisfying $\delta_1 < \ldots < \delta_C.$ Then, as in AM, it follows from previous work \cite{montesinos2015genomic,sorensen1995bayesian} that \begin{equation*} \pi(\delta_c | \boldsymbol{y}, \boldsymbol{z}) \propto \frac{1}{\min(z_i|y_i = c+1)-\max(z_i|y_i=c)}\mathbb{I}(\boldsymbol{\delta} \in T), \end{equation*} which leads to \begin{equation*} \delta_c | \boldsymbol{y}, \boldsymbol{z} \sim \mathcal{U}(\min \{ \max(z_i | y_i = c), \delta_{c+1}, \delta_C \}, \max \{ \min(z_i | y_i = c + 1), \delta_{c-1}, \delta_0 \} ). \end{equation*}

\section{Supplementary: Analysis for Early Puberty Application with Unnormalized Covariates}

\begin{figure}[!ht]
\centering
\includegraphics[width=0.49\textwidth]{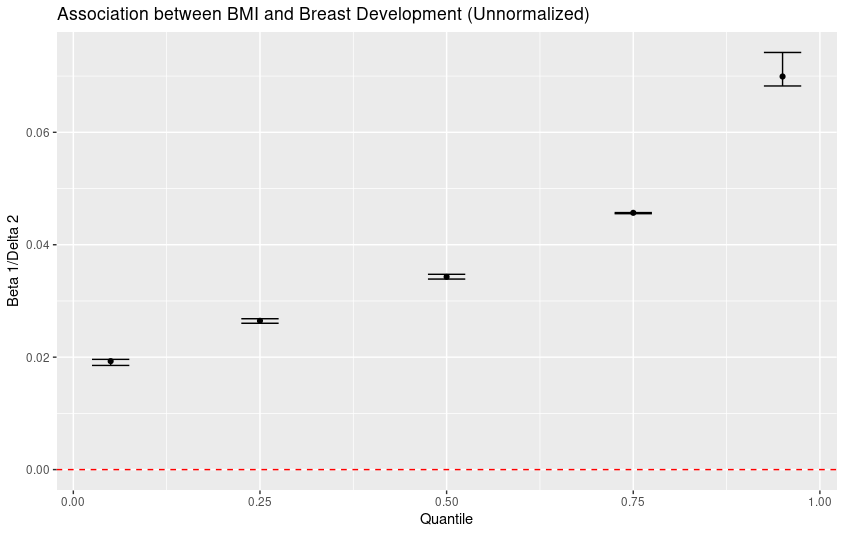}
\includegraphics[width=0.49\textwidth]{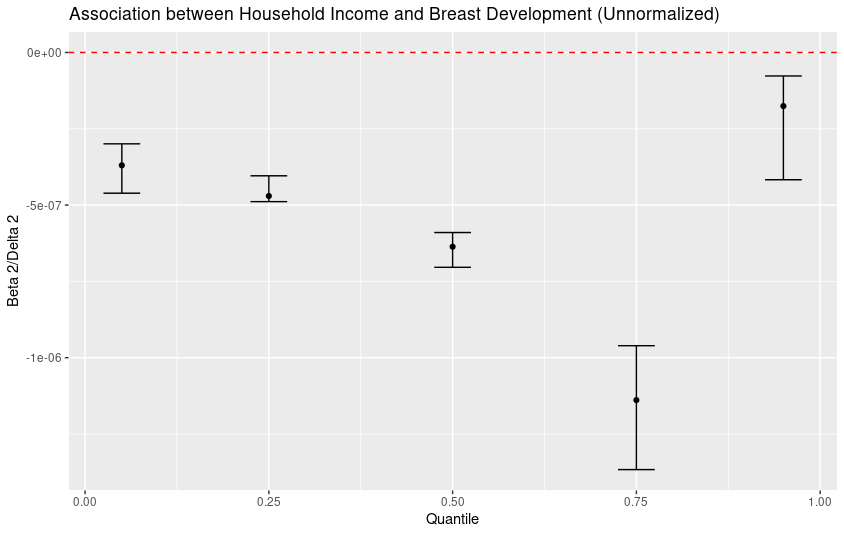}
\caption{Coefficients found by BORPS across quantiles for BMI (left) and income (right) for breast development in girls. The x-axis represents five quantiles, and the y-axis represents the ratio $\frac{\beta}{\delta_2}$ for the appropriate coefficient, with the points representing posterior mean estimates and whiskers representing 95\% confidence intervals. The red dashed line represents the 0 line.} 
\label{fig:bmi-income}
\end{figure}

\begin{figure}[!ht]
\centering
\includegraphics[width=0.49\textwidth]{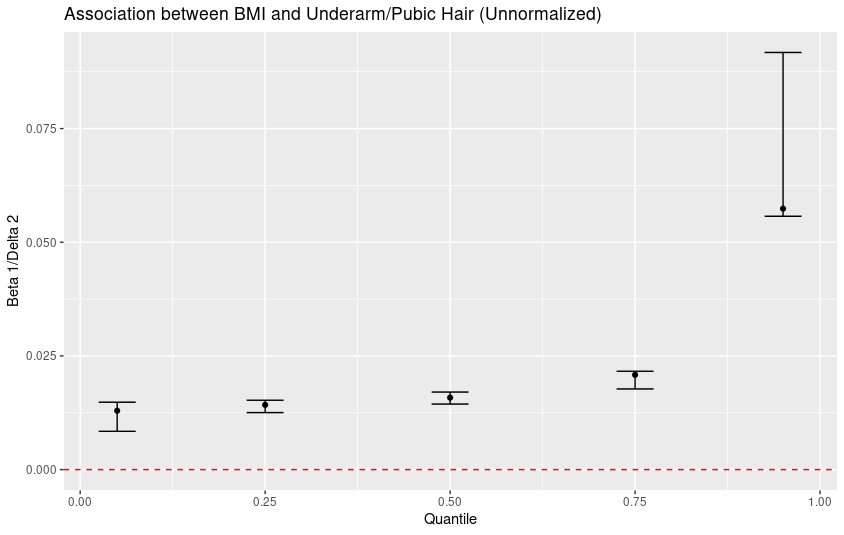}
\includegraphics[width=0.49\textwidth]{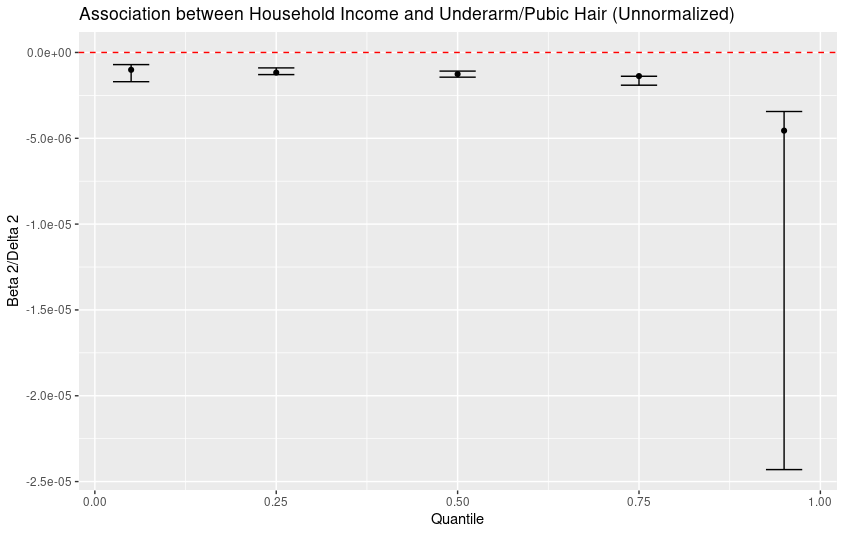} 
\caption{Coefficients found by BORPS across quantiles for BMI (left) and income (right) for underarm/pubic hair in boys. The x-axis represents five quantiles, and the y-axis represents the ratio $\frac{\beta}{\delta_2}$ for the appropriate coefficient, with the points representing posterior mean estimates and whiskers representing 95\% confidence intervals. The red dashed line represents the 0 line.} 
\label{fig:bmi-income-boys}
\end{figure}

Above are the results of the early puberty application using unnormalized, rather than z-scored, covariates. The overall trend is the same as was observed with the z-scored covariates, except for the sign change at the most extreme (0.95th) quantile.

\end{document}